\documentclass[a4paper,11pt]{article}
\usepackage{pos}
\usepackage{graphicx}
\usepackage{subcaption}
\usepackage{soul}

\title{A two-zone emission model for Blazars and the role of Accretion Disk MHD winds}
 \ShortTitle{2-zone Blazar emission and their classification}
\author*[a]{Stella Boula}
\author[a]{Apostolos Mastichiadis}
\author[b]{Demosthenes Kazanas}

\affiliation[a]{Department of Physics, National and Kapodistrian University of Athens, Panepistimiopolis, GR 15783 Zografos, Greece}
\affiliation[b]{NASA Goddard Space Flight Center, Greenbelt, MD, United States}

\emailAdd{stboula@phys.uoa.gr}

\abstract{
Blazars are a sub-category of radio-loud active galactic nuclei with relativistic jets pointing towards the observer. They exhibit non-thermal variable emission, which practically extends over the whole electromagnetic spectrum. Despite the plethora of multi-wavelength observations, the origin of the emission in blazar jets remains an open question. In this work, we construct a two-zone leptonic model: particles accelerate in a small region and lose energy through synchrotron radiation and inverse Compton Scattering. Consequently, the relativistic electrons escape to a larger area where the ambient photon field, which is related to Accretion Disk MHD Winds, could play a central role in the gamma-ray emission. This model explains the Blazar Sequence and the broader properties of blazars, as determined by Fermi observations, by varying only one parameter, the mass accretion rate onto the central black hole. Flat Spectrum Radio Quasars have a strong ambient photon fields and their gamma-ray emission is dominated by the more extensive zone, while in the case of BL Lac objects, the negligible ambient photons make the smaller, i.e. acceleration, zone dominant.}
\FullConference{37$^{\rm{th}}$ International Cosmic Ray Conference (ICRC 2021)\\
		July 12th -- 23rd, 2021\\
		Online -- Berlin, Germany}

\begin{document}
\maketitle

\section{Introduction}
 Blazars (Flat Spectrum Radio Quasars and BL Lac objects) are the most extreme type of Active Galactic Nuclei (AGN), with
relativistic jets pointing toward us. Their emission is non-thermal and it is amplified due to the relativistic bulk motion of the jets.
Blazar Spectral Energy Distribution (SED) consists of two broad ``humps'', one spanning from radio to optical-UV (and occasionally X-ray) bands and another one
that extends from X-rays to multi-GeV and occasionally to TeV $\gamma$-rays. The low-frequency component is believed to be due to synchrotron emission by non-thermal electrons, while the higher one is attributed to inverse Compton (IC) scattering of the relativistic electrons on
synchrotron or external photons. Relativistic plasma motion and radiative cooling affect the temporal variability of these sources, which helps define the emitting region's size. 
 Certain patterns in the spectral features became apparent and are now known as the ``Blazar
Sequence" \cite{Foss98}. Blazars become redder with increasing bolometric luminosity
$L_{\rm bol}$, in that their synchrotron peak frequency $\nu_{\rm pk}^{\rm syn}$ decreases as
$L_{\rm bol}$ increases; at the same time, their Compton dominance (CD; i.e., the ratio of their IC to synchrotron luminosities) increases and so do their $\gamma$-ray spectral indices, i.e. their spectral index becomes steeper.
The Blazar Sequence, established originally with 132 objects out of which only 33 were detected in high
energy $\gamma$-rays, was supplemented with the launch of \emph{Fermi} and the discovery
of more than 5000 $\gamma$-ray Blazars as recorded in the  4th Fermi Blazar Catalog
 \cite{4th}. This
compilation provided novel correlations that replaced those of the original Blazar Sequence. This result is implying that the underlying physics is probably related to fundamental parameters of the AGN phenomenology. \\
\indent The optically thin Blazar GeV emission suggests its location to be far from the
accreting black hole (BH), possibly out to a distance as large as  $10^6R_S~\sim$  10 pc \cite{Marscher10} (where $R_S$ is the Schwarzschild radius ). Furthermore, the AGN torii (dusty, molecular structures of height/radius ratios
$z/D \simeq 1$)  invoked in the unification of the radio-quiet or radio-loud AGN subclasses \cite{AntonMill}, are of similar scales and, as we argue, play a significant role in Blazar physics. To reconcile the discrepancy of the torii geometry expected, \cite{KK94} proposed that these torii are, in fact, MHD accretion disk winds \cite{BKM19, CL94}; these are launched across the entire disk, from the BH
vicinity of a few $R_S$ to the BH influence radius $D$  $\sim~(c/\sigma)^2 R_S \sim 10^6 R_S \sim 10 pc$ (for $M_{BH}\simeq 10^8M_{\odot}$). Furthermore, the discovery of Warm Absorbers (WA, blue-shifted absorption features) and their successful modeling as photoionized MHD
winds that extend to $r \sim 10^6R_S$ \cite{Behar09,FKCB}, established the combined AGN WA-torii as a
single entity.
Finally, modeling the absorbers of the Galactic BH GRO 1655-40 with the same type of winds \cite{FKCB17}
suggests the possibility of their presence in any accreting BH. \\
\indent The physical properties of such MHD winds depend on a few parameters and their presence allowed us to reproduce a theoretical Blazar Sequence by varying only one parameter, namely the mass accretion rate, \cite{BKM19,20BMK}. Here we will show the results of a two-zone emission model based on the extension of our previous works. However, now we assume that electrons are accelerated into a zone close to the central engine and escape and radiate into a larger region, which will refer as the cooling zone.
\vspace{-0.2cm}
\section{Emission model-Scalings to mass accretion rate}

\indent The broader morphology of the non-thermal Blazar SED depends
on the ratio of the magnetic to photon energy densities. To calculate the former we assume some sort of equipartition with the accreting matter, the energy density. By assuming the power of the accretion $P_{\rm acc}= \dot{m}{\cal M}L_{\rm Edd}$, with  $\dot{m}$ the mass accretion rate normalized to the Eddington one and 
${\cal M}= M_{\rm BH}/M_{\rm \odot}$, where
$M_{\rm BH}$ is the mass of the black hole,  one can calculate the magnetic field energy density at position z:
%Here we assume that the external photon field is related to photons which are scattered on accretion disk wind particles \cite{BKM19}. To calculate the Blazar SED, we solve the kinetic equations of electrons and photons as described in \cite{MK95}. We assume $P_{\rm acc}= \dot{m}{\cal M}L_{\rm Edd}$, with  $\dot{m}$ the mass accretion rate normalized to the Eddington one and 
%${\cal M}= M_{\rm BH}/M_{\rm \odot}$ where
%$M_{\rm BH}$ is the mass of the black hole. 
%Assuming equipartition the magnetic filed and the accreting matter energy density one can write:
\begin{equation}\label{B}
 U_{\rm B} \propto \eta_{\rm b} \dot{m}{\cal{M}}^{-1},
\end{equation}
%where $U_B$ is the magnetic field energy density.
where $\eta_{\rm b}$ is a proportionality constant.
We assume for the external photon field that is related to photons which are scattered on accretion disk wind particles \cite{BKM19} and thus the external photon field density $U_{ext}$ in the jet frame has the form: 
 \begin{equation}\label{ext}
  U_{\rm ext}= \Gamma^2 {\rm U}_{\rm sc}\propto \Gamma^2 \epsilon \dot{m}^{\alpha+1}{\cal M}^{-1}~~(\alpha =1 ~{\rm{for}}~\dot{m}\geq 0.1 ~~{\rm{and}}~ \alpha=2 ~{\rm{for}}~\dot{m}<0.1),
 \end{equation}
 where $\epsilon$ is  the  efficiency  of  the  conversion
of  the  accreting  mass  into  radiation and  $\Gamma$ is the bulk Lorentz factor of the source. We have assumed that the disk emits like a black body characterized by a temperature $T_{\rm disk} $ in order to estimate the spectrum of the scattered photons. As we pointed in \cite{BKM19}, all  input parameters  required for the calculation of the spectrum are scaled with $\dot{m}$ and ${\cal{M}}$. Using the above definitions for the physical properties of the source, we obtain the Blazar SED, by solving the coupled integro-differential kinetic equations of electrons and photons as described in \cite{MK95}.
 We emphasize that according to relations \ref{B},\ref{ext} the basic parameters of the system of equations depends basically only on the mass accretion rate, $\dot{m}$.

 Therefore, we assume a blob of plasma of radius $R$ where particles are accelerated with a characteristic timescale $t_{acc}$, (e.g. \cite{KRM98}). In the case of the first order Fermi acceleration we have \begin{equation}\label{tacc}
    t_{{acc}_{FI}}\geq 6\left(\frac{c}{u_s}\right)^2\frac{\lambda}{c}\simeq 6 \frac{r_g c}{u_s^2},
\end{equation}
where 
\begin{equation}
    r_g=\frac{\gamma m c^2}{eB}.
\end{equation}

\begin{figure}[!htbp] % "[t!]" placement specifier just for this example
\begin{subfigure}{0.6\textwidth}
\includegraphics[width=\linewidth]{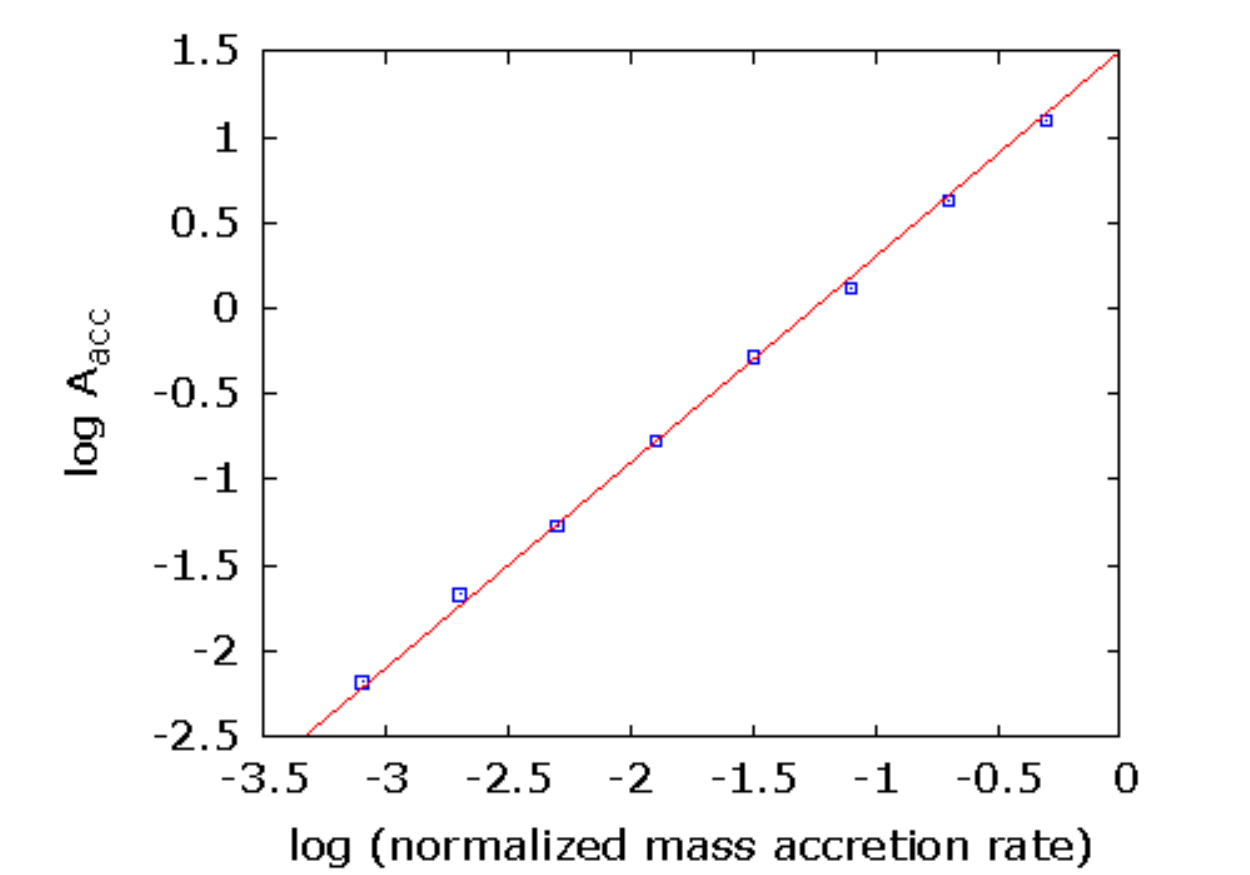}
 \label{fig:a}
 \end{subfigure}\hspace*{\fill}
\begin{subfigure}{0.5\textwidth}
\includegraphics[width=\linewidth]{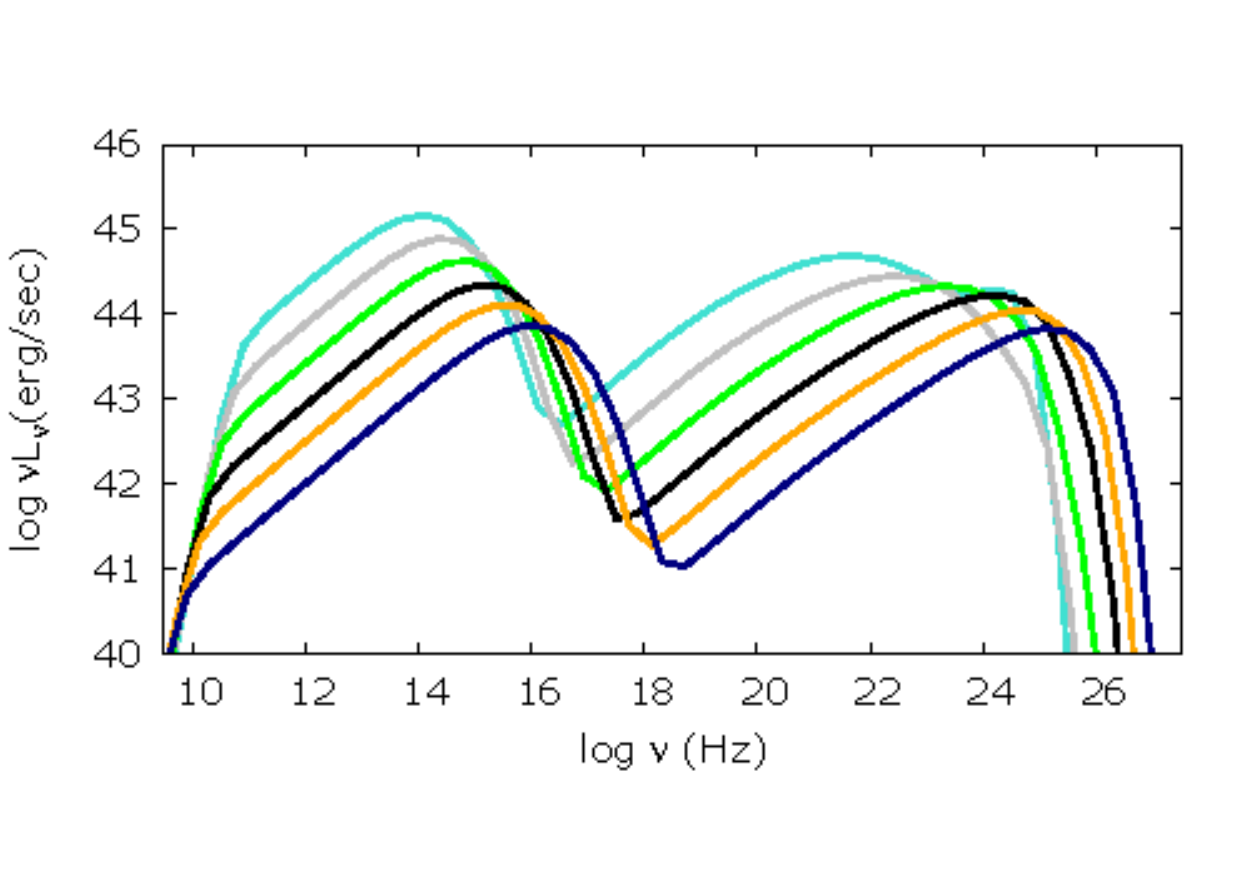}
\label{fig:b}
\end{subfigure}
\caption{\textbf{Left:} The dependece of the acceleration timescale to the normalized mass accretion rate. \textbf{Right:} The calculated Spectral Energy Distribution of BL Lac objects in the case for various values of the mass accretion rate.  } \label{fig:1a}
\end{figure}
In our study, we re-write the Equation \Ref{tacc} as $t_{acc}\propto \gamma A_{acc}(\dot{m})$ taking into account the dependence on the mass accretion rate. By solving the set of equations we find there is an almost linear dependence between $t_{acc}$ and $\dot{m}$ in order to explain the high peaked synchrotron blazars, see Figure~\ref{fig:1a}.

\section{Particles escape: A two-zone emission model}
\begin{figure}[!htbp]
    \centering
    \includegraphics[width=0.75\linewidth]{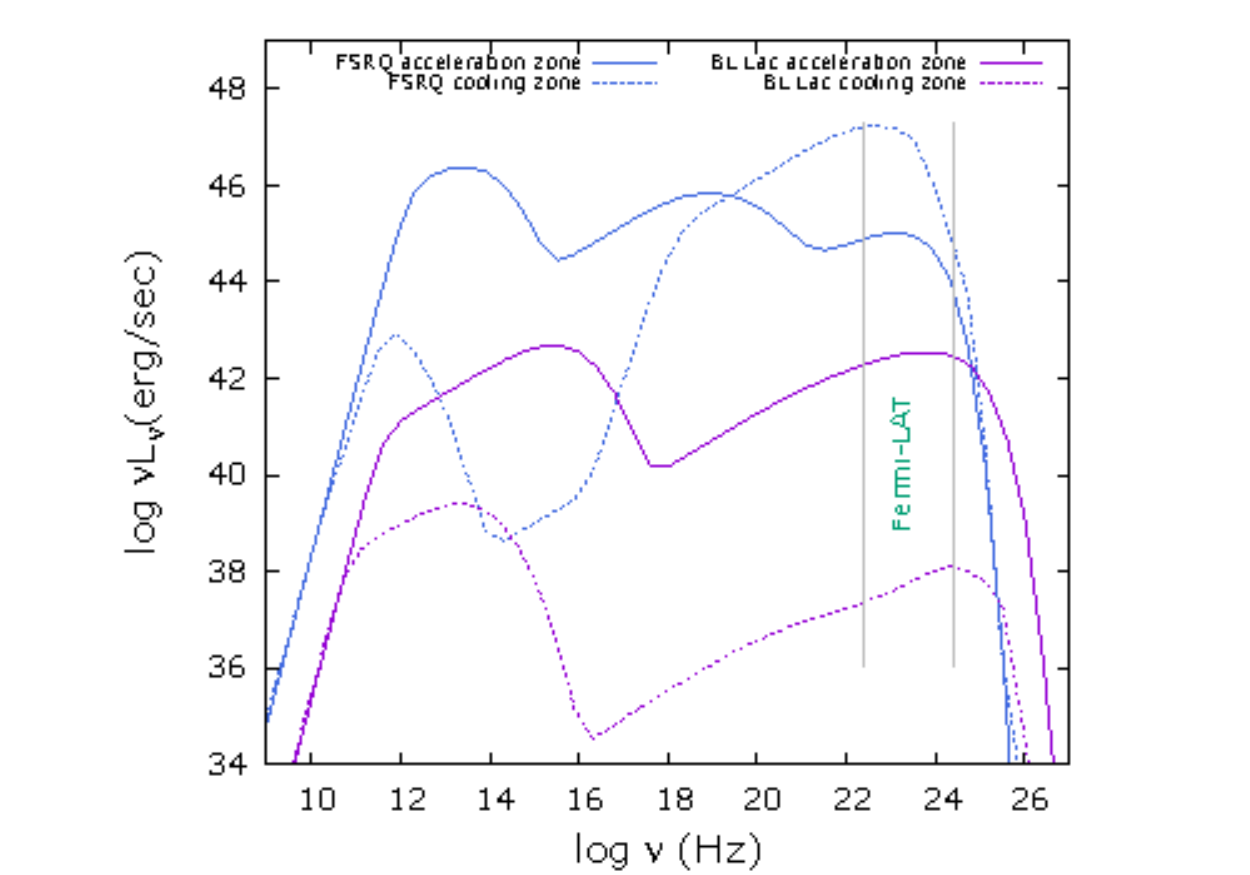}
    \caption{Results for FSRQ and BL Lac objects according to the two-zone model. Straight lines depict the emission from the acceleration zone and dotted lines the emission from the cooling zone.}
    \label{fig:sed_2}
\end{figure}

As a second step, we study a two zone model: particles  accelerate in a zone closer to the central engine, then they escape to a larger volume further away and cool due to synchrotron and inverse Compton losses. 
\begin{figure}[!htbp]
    \centering
    \includegraphics[width=0.5\linewidth]{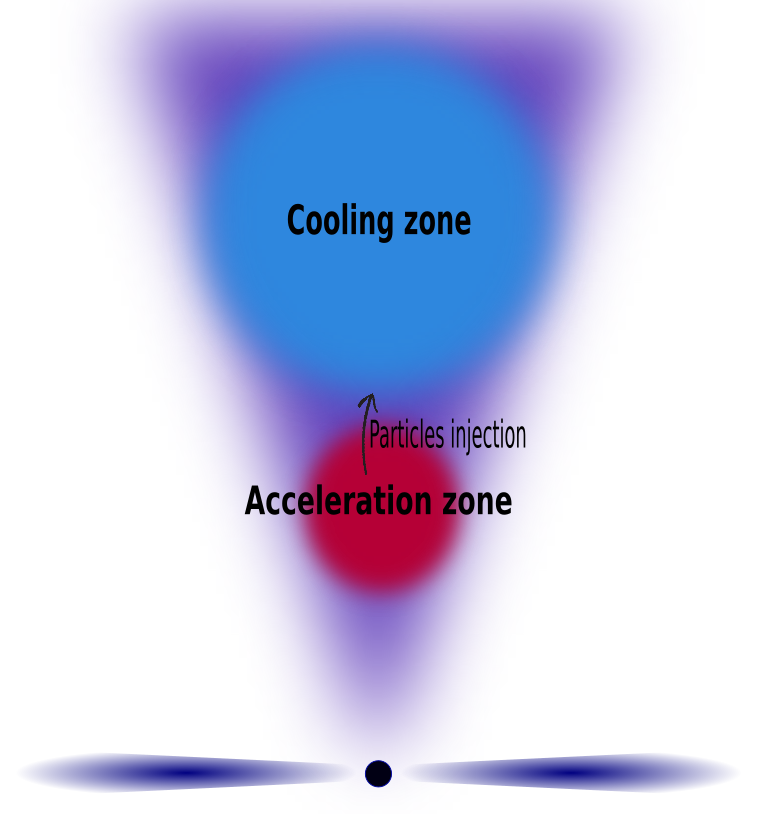}
    \caption{Sketch of the two-zone model. Particles accelerate and radiate in the acceleration zone (red). While, a fraction of the accelerated particles escapes into a larger volume(blue), where they cool and radiate.}
    \label{fig:model_2zone}
\end{figure}
The kinetic equation of electrons in the first zone (or acceleration zone) has the form: 
 \begin{equation}
     \frac{\partial n_{e_I}(\gamma,t)}{\partial t}+\frac{n_{e_I}(\gamma,t)}{t_{{esc}_I}(\dot{m},\gamma)}+\frac{\partial}{\partial \gamma}\left[\frac{\gamma}{t_{acc}(\dot{m},\gamma)}n_{e_I}(\gamma,t)\right]=\mathcal{L}_{syn}(\gamma,t) + \mathcal{L}_{ICS}(\gamma,t).
 \end{equation}
The kinetic equation of relativistic electrons in the second zone (or cooling zone) has the form: 
 \begin{equation}
     \frac{\partial n_{e_{II}}(\gamma,t)}{\partial t}+\frac{n_{e_{II}}(\gamma,t)}{t_{{esc}_{II}}}=Q_{inj}+\mathcal{L}_{syn}(\gamma,t) + \mathcal{L}_{ICS}(\gamma,t),
 \end{equation}
where $n_{e_I}, ~n_{e_{II}}$ the electron differential distribution function in the first and second zone respectively, $t_{{esc}_I},~t_{{esc}_{II}}$ the electron escape timescale from the first and the second zone respectively, the term $Q_{inj}=\frac{n_{e_I}}{tesc_I}$ refers to the relativistic electrons of the acceleration zone that escape and inject to the cooling zone, (e.g. \cite{1998KRM}). 
To calculate synchrotron losses, we assume that the magnetic field decreases with the distance $z$ from the central engine as $B\propto 1/z$. Furthermore, the energy density of the external photon field $U_{ext}$ is constant with the distance z, \cite{BKM19}. According to our assumption, the two classes of blazars show a very distinct behavior: 
\begin{itemize}
\item BL Lac objects emission is dominated by the first zone. This results from the fact that the energy density of the magnetic field dominates in both zones. The reason is that in the acceleration zone, we have $U_{B_I}>>U_{ext}$ as BL Lac objects have a weak external photon field. In the radiation zone\footnote{The size of the second zone is related to the electrons cooling timescale, in order the losses to be Compton dominated.} even the fact that the magnetic energy density $U_B$ decreases with z (see above), 
we still have that $U_{B_{II}} >> U_{ext}$ everywhere, where $U_{B_{II}}$ the magnetic field strength in the second zone, see Figure \Ref{fig:sed_2}. As a result, the spectrum is dominated by the emission of the acceleration zone. 
\item FSRQs synchrotrom emission is mainly produced by the first zone, while the second zone is dominated by the inverse Compton scattering. For the parameter set that we study in the acceleration zone, the magnetic energy density is again more significant than the energy density of the external photon field. However, the second zone is further from the central engine, where the magnetic field decreases, but the external photons’ energy density remains constant \cite{BKM19}. Now we have $U_{ext}>>U_{B_{II}}$ and the contribution of the external inverse Compton scattering is more significant than this of the first zone. As a result, in the total flux, the low-frequency component is related to the synchrotron radiation from the acceleration zone and the high one to the inverse Compton scattering losses from the cooling zone. 
\end{itemize}
The above can be illustrated by Figure \Ref{fig:sed_2} which depicts our calculations for the two zones in the cases of FSRQs and BL Lac objects. 
\begin{figure}[!htbp]
    \centering
    \includegraphics[width=0.75\linewidth]{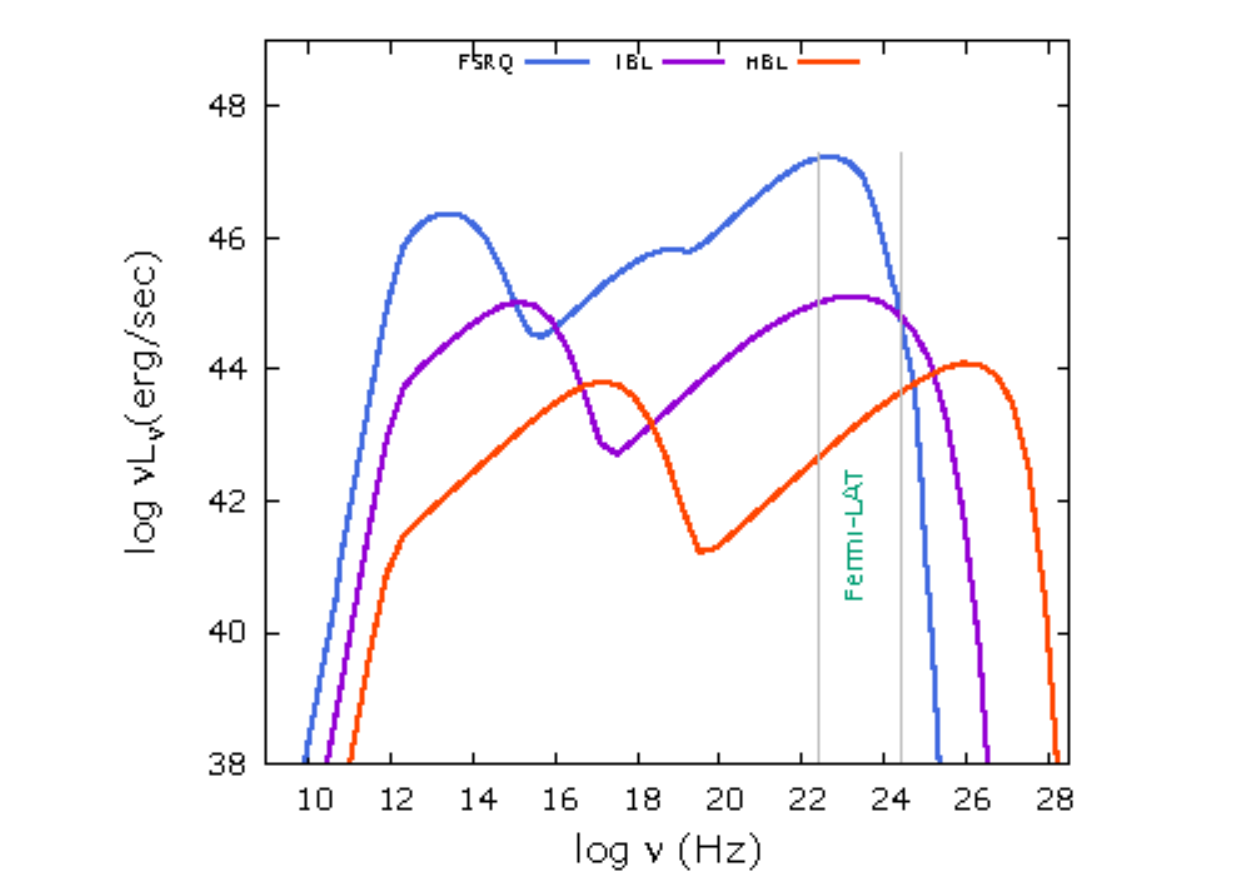}
    \caption{The theoretical Blazar Sequence according to the superposition of the two-zone emission by varying only the mass accretion rate, see Table \Ref{tab3} for the values of the input parameters. The acceleration zone is at a distance of $z=0.01~pc$ from the central black hole which is assumed to have  a mass $\mathcal{M}=10^8$. The external photon field is produced from the isotropic scattering of disk photons on the wind particles between radii $R_1=9\cdot 10^{14}~\rm{cm}$ and $R_2=3\cdot 10^{18}~\rm{cm}$. The efficiencies of the magnetic field and the external photon field are $\eta_{\rm b}=0.01$, and $\epsilon=0.05$, respectively. The number of the injected electrons in the acceleration process, which is assumed to be of Type I Fermi, depends linearly on the mass accretion rate. The bulk Lorentz factor is $\Gamma=30$ and the Doppler factor is $\delta=15$. The characteristic temperature of the accretion disk is $T_{disk}=3 \cdot 10^3~$K.}
    \label{fig:sed_2zone}
\end{figure}
\begin{figure}[!htbp]
    \centering
    \includegraphics[width=0.75\linewidth]{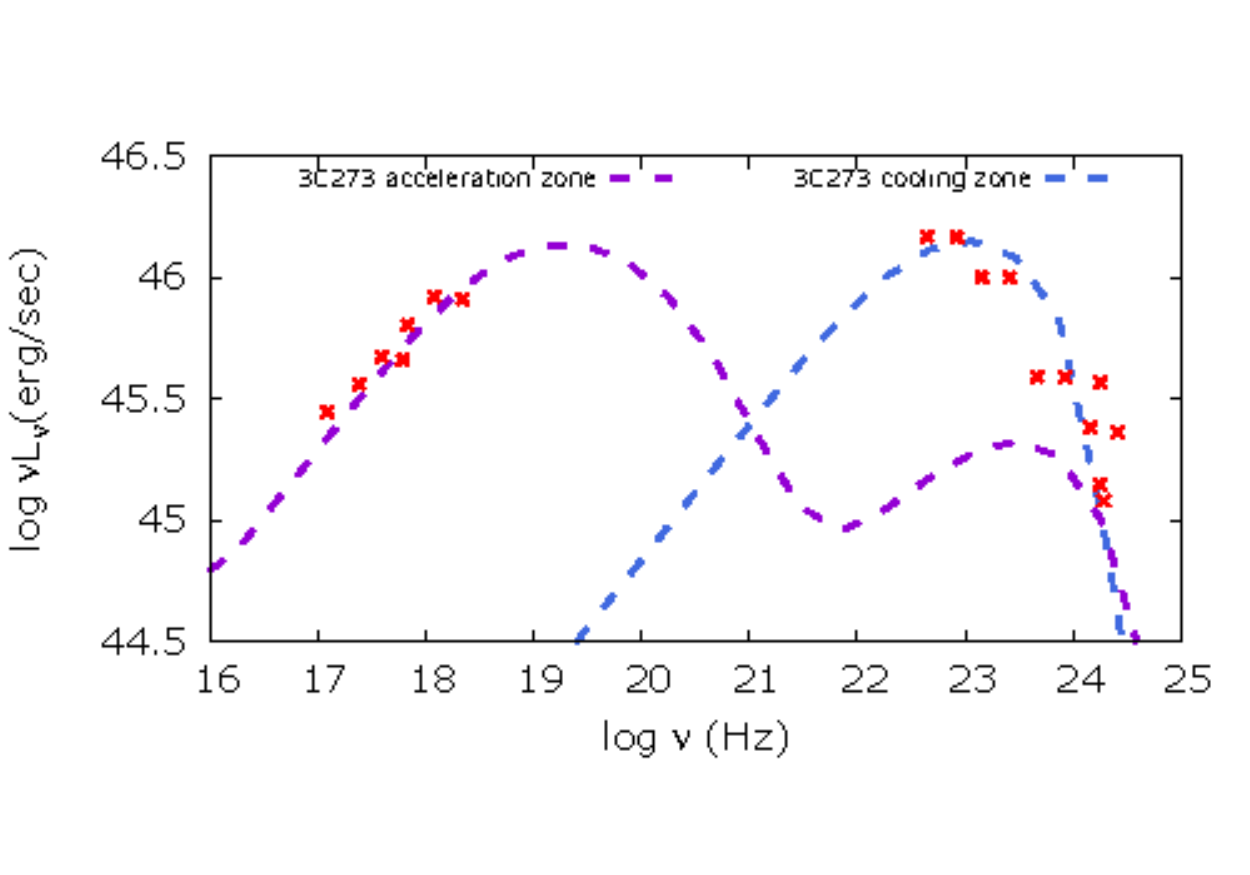}
    \caption{An application of the two-zone model in the case of FSRQ 3C273. The acceleration zone is at distance $z=0.01pc$.  The external photon field is produced from the isotropic scattering of disk photons on the wind particles between radii   $R_1=9\cdot 10^{14}~\rm{cm}$ and $R_2=3\cdot 10^{18}~\rm{cm}$.  The magnetic field strength in the acceleration zone is $B=1~G$, its radius is $R=5\cdot 10^{15}$ cm, and the energy density of the external photon field is $U_{ext}=2.5\cdot10^{-3} ~\frac{erg}{sec}$. The bulk Lorentz factor is $\Gamma=30$ and the Doppler factor is $\delta=15$. The characteristic temparature of the accretion disk is $T_{\rm disk}=3 \cdot 10^4$ K. Data are reproduced from \cite{GIommi12P}.}
    \label{fig:3c273}
\end{figure}
\begin{table} 
\begin{center}
%\resizebox{\textwidth}{!}
{
  \begin{tabular}{ |c|c | c | c | c |}
    \hline
    $ \dot{m}$ & $ B({\rm G}) $ & $U_{\rm ext}$ $\left(\frac{\rm erg}{\rm{cm}^3}\right)$ & $A_{\rm acc}$  & Blazar Class\\ \hline
         -0.5 & 1  & -2.6 &-4& FSRQ\\ \hline
     -1.5 & 0 & -5.6 & -5& LBL \\ \hline
    -2.5 & -1  & -8.6 &-6&  HBL \\    
    \hline
  \end{tabular}} 
\end{center}
\caption{The values of the  input parameters for different mass accretion rates when particle acceleration is included in the numerical code. All the values are in a logarithmic scale. }\label{tab3}
\end{table} 

In Figure \Ref{fig:sed_2zone} we present our results for a theoretical Blazar Sequence in the case of the two-zone model, see Table \ref{tab3} for the values of the input parameters and the dependence of $\dot{m}$. In both classes of blazars particles escape from the acceleration zone. However, FSRQs present a  characteristic signature in $\gamma$-rays, because particles inject into the cooling zone and interact with a strong external photon field. On the contrary, in BL Lac objects, when particles escape, they do not interact with a strong external photon field and as a result, the cooling zone has a lower contribution to the total flux. 

In Figure \Ref{fig:3c273} we show our results in applying a two-zone model in the case of 3C273, an FSRQ object. We zoom on the high-energy part of the spectrum (X-rays, $\gamma$-rays). X-rays are produced by SSC of the acceleration zone and $\gamma$-rays by the external Compton of the cooling zone.

\section{Conclusion}
In this work, we reproduce the theoretical Blazar Sequence based on the model of \cite{BKM19} by varying only the mass accretion rate that seems to explain the Blazar phenomenology. We solve self-consistently the electron and photon kinetic equations, by assuming that electrons accelerate into a small region and lose energy through synchrotron and inverse Compton scattering. While, a part escapes to a larger volume where they lose energy through the same physical processes. Under this assumption, we produce the theoretical Blazar Sequence by adding up the fluxes of the two zones in both the cases of FSRQs and BL Lac objects. 

\bibliographystyle{JHEP}
\bibliography{new}

\end{document}